\def\cm2{cm$^{-2}$}
\def\cm3{cm$^{-3}$}

\def\radec{\hbox{RA, Dec.(J2000)}}
\def\gra{$^{\circ}$}

\documentclass[twocolumn,tighten]{aastex61}

\usepackage{natbib}

\newcommand\aastex{AAS\TeX}

\received{}
\revised{}
\accepted{}
\submitjournal{The Astrophysical Journal}

\shorttitle{\aastex\ }
\shortauthors{Mendoza et al.}

\begin{document}

\title{G331.512-0.103: An Interstellar laboratory for molecular synthesis\\ I The ortho-to-para ratios for CH$_3$OH and CH$_3$CN}

\correspondingauthor{Edgar Mendoza}
\email{emendoza@usp.br}

\author{Edgar Mendoza}
\affil{Universidade de S\~ao Paulo, IAG
Rua do Mat\~ao, 1226, Cidade Universit\'aria, 05508-090, S\~ao Paulo, Brazil}

\author{Leonardo Bronfman}
\affiliation{Departamento de Astronom{\'{\i}}a, Universidad de Chile, Casilla 36-D, Santiago de Chile, Chile}

\author{Nicolas U. Duronea}
\affil{Instituto Argentino de Radioastronom{\'{\i}}a, IAR 
CONICET, CCT-La Plata, C.C.5., 1894, Villa Elisa, Argentina}

\author{Jacques R. D. L\'epine}
\affil{Universidade de S\~ao Paulo, IAG 
Rua do Mat\~ao, 1226, Cidade Universit\'aria, 05508-090, S\~ao Paulo, Brazil}

\author{Ricardo Finger}
\affiliation{Departamento de Astronom{\'{\i}}a, Universidad de Chile, Casilla 36-D, Santiago de Chile, Chile}

\author{Manuel Merello}
\affiliation{Istituto di Astrofisica e Planetologia Spaziali-INAF
Via Fosso del Cavaliere 100, I-00133 Roma, Italy}

\author{Carlos Herv\'ias-Caimapo}
\affiliation{Departamento de Astronom{\'{\i}}a, Universidad de Chile, Casilla 36-D, Santiago de Chile, Chile}
\affil{Jodrell Bank Centre for Astrophysics, School of Physics and Astronomy, University of Manchester, Oxford Road, Manchester M139PL, UK}

\author{Diana R. G. Gama}
\affil{Universidade de S\~ao Paulo, IAG
Rua do Mat\~ao, 1226, Cidade Universit\'aria, 05508-090, S\~ao Paulo, Brazil}

\author{Nicolas Reyes}
\affiliation{Electrical Engineering Department, Universidad de Chile, Av. Tupper 2007, Santiago, Chile}

\author{Lars \r{A}ke-Nyman}
\affiliation{Joint ALMA Observatory, JAO 
Alonso de C\'ordova 31070, Vitacura,  Santiago de Chile, Chile}



\begin{abstract}

Spectral line surveys reveal rich molecular reservoirs in G331.512-0.103, a compact radio source in the center of an energetic molecular outflow. In this first work, we analyse  the physical conditions of the source by means of CH$_3$OH and CH$_3$CN.\\ 
The observations were performed with the APEX telescope. Six different system configurations were defined to cover most of the band within (292--356)~GHz; as a consequence we detected a forest of lines towards the central core.\\ 
A total of 70 lines of $A/E$-CH$_3$OH and $A/E$-CH$_3$CN were analysed, including torsionally excited transitions of CH$_3$OH~($\nu_t$=1). In a search for all the isotopologues, we identified transitions of $^{13}$CH$_3$OH.  The physical conditions were derived considering collisional and radiative processes. We found common temperatures for each $A$ and $E$ symmetry of CH$_3$OH and CH$_3$CN; the derived column densities indicate an $A/E$ equilibrated ratio for both tracers.\\
The results reveal that CH$_3$CN and CH$_3$OH trace a hot and cold component with $T_k \sim$~141~K and $T_k \sim$~74~K, respectively. In agreement with previous ALMA observations, the models show that the emission region is compact ($\lesssim5.5\arcsec$) with gas density $n$(H$_2$)=(0.7--1) $\times$ 10$^7$~cm$^{-3}$. The CH$_3$OH/CH$_3$CN abundance ratio and the evidences for pre-biotic and complex organic molecules suggest a rich and active chemistry towards G331.512-0.103.
\end{abstract}
%
%
%
\section{Introduction} 
\label{sec:intro}

G331.512-0.103 is one of the most luminous and energetic molecular outflows known in the Galaxy. All the system is embedded in a star forming region known as G331.5-0.1, located in the Norma spiral arm at a distance of $\sim$~7~pc. This source exhibits typical H~II region properties and intense maser emission of OH and CH$_3$OH \citep{Caswell1998,Nyman2001,bro08}.
At the center of the system, a young and massive stellar object drives a bipolar flow of around 55$M_{\odot}$ with a momentum of $\sim$~2.4 $\times$~10$^3$~$M_{\odot}$~km~s$^{-1}$. 
Radio continuum observations have revealed a compact and central structure associated to a dust core. The central region has been also mapped with ALMA observations with transitions of SiO (8--7), H$^{13}$CO$^+$ (4--3), HCO$^+$ (4--3) and CO (3--2). The observations revealed a ring-like structure, consistent with a cavity, and the existence of a high velocity outflow emission confined in a region lower than~5$\arcsec$ in size \citep{mer13b,mer13a,Hervias2015}. In this work, we present new results on the emission of CH$_3$OH and CH$_3$CN detected towards the central core, denominated hereafter as G331. 

As an interstellar laboratory, G331 exhibits active chemistry in pre-biotic and complex organic molecules likely stimulated by physical processes in the molecular core and outflow. With the Atacama Pathfinder EXperiment APEX telescope, we have detected a vast number of molecular lines potentially linked to reservoirs traced by CH$_3$OH and CH$_3$CN. Those tracers are excellent candidates to unveil the different physical components that O- and N-bearing molecules can reveal towards hot molecular cores 
\citep{sand94,Wyrowski1999,jor04,Beuther2005,Fontani2007,Fuente2014,giannetti2017}. In addition, from a chemical point of view, CH$_3$OH and CH$_3$CN are considered as parent and daughter molecules, respectively, in the route of interstellar molecular formation (e.g. \citealt{Nomura2004}). With that motivation, we started a systematic study of G331 beginning with spectral analyses of $A/E$-CH$_3$OH and $A/E$-CH$_3$CN.

\subsection{The tracers CH$_3$CN and CH$_3$OH} 
\label{intro1.1}

CH$_3$CN (methyl cyanide) is a symmetric top molecule. The nuclear spin state of the hydrogen atoms  defines whether the molecule has an ortho ($E$) or  para ($A$) symmetry. Rotational levels of CH$_3$CN are characterized by two quantum numbers: the total angular momentum ($J$) and its projection on the axis of symmetry ($K$). Spectral signatures are associated to $K$-ladder structures whose notation for the $A$ and $E$ configurations are $K=3n$ and $K=3n\pm$1, respectively \citep{Solomon1971,bou80,Cummins1983}.

CH$_3$OH (methanol) is one of the simplest asymmetric-top molecules, however its spectrum is quite complicated since there is a strong coupling between torsional and vibrational modes. Methanol also exists in two species denoted as $A$ and $E$ depending on the nuclear spin alignment of the three H-atoms of the methyl group (CH$_3$). The energy levels of methanol can be assigned with the total angular momentum $J$ and its component $K$ along the symmetry axis \citep{lees68,Lovas82}. 

The excellent spectral resolution of the used instrument allowed us to study separately the $A$ and $E$ isomers of CH$_3$CN and CH$_3$OH. Nuclear spin conversion of $A$ and $E$ symmetries are considered as rare events, being affected by chemical reactions, non-reactive collisions and grain-surface mechanisms  \citep{Willacy1993,Hugo2009}.  However, the spin conversion has a particular relevance in molecular astrophysics. Important questions remain unsolved, for instance, how collisions stimulate conversions and, as a consequence, if that property can be used as an astronomical clock \citep{Lee2006,Sun2015}. One of our goals is to verify if the $A$ and $E$ pairs are equally populated at the local temperature of the source (e.g \citealt{Andersson1984,Minh1993,Wirstrom2011}).

We organized this paper as follows: in section~\ref{sec2} the observational procedure is described. In section~\ref{sec3} results about the line identification are presented. In section~\ref{sec4} we discuss the radiative analysis. The discussion, conclusions and perspectives are presented in sections~\ref{sec5}~and~\ref{sec6}. 

\section{Observations and methodology}
\label{sec2}

The observations were carried out in March 2016 with the Atacama Pathfinder Experiment Telescope (APEX), located at {\it Llano de Chajnantor} (Chilean Andes). The spectra were obtained using the single point mode towards the coordinates  \radec\ = 16$^h$12$^m$10.1$^s$, $-$51\gra28\arcmin38.1\arcsec. We used the APEX-2 receiver of the Swedish Heterodyne Facility Instrument (SHeFI) as frontend \citep{gu06,ris06}. The employed backend  was the eXtended bandwidth Fast Fourier Transform Spectrometer2 (XFFTS2),  which consists of two units with a bandwidth of 2.5~GHz divided into 32768 channels. We selected six set-ups to cover a band ranging from 292~GHz to 356~GHz; they are summarized in Table~\ref{tab1}.  The integration time  was estimated to obtain a conservative rms noise of $\sim$~25 mK; then each set-up expended integration times between 0.7 and 2.6~hs, with system temperatures in the range of $T_{sys}\sim$(200--380)~K.\\ 
The calibration was done applying the chopper-wheel technique. The output intensity provided by the system was obtained in a scale of $T_{\rm A}^*$, which represents the antenna temperature corrected by atmospheric attenuation. The observed intensities were converted to the main-beam temperature scale using $\eta_{\rm mb}$ = 0.73, the main beam efficiency for APEX-2  \citep{gu06}.  

All the $K$-ladders structures appeared closely spaced in a wide range of frequencies, so that they were observed with the same receiver. This fact helped us with the inspection of irregularities due to calibration uncertainties, for which we adopted an error of 20~\%.
\begin{table}[t!]
\centering
\caption{List of set-ups observed with the APEX-2 SHeFI instrument towards G331. \label{tab1}}
\begin{tabular}{cccc}
\hline
\hline
Set-up      & Frequency   &  Beam        &   Resolution                \\        
centred at  &      (GHz)  &  ($\arcsec$) &   (10$^{-2}$ km s$^{-1}$)     \\ 
\hline
CH$_3$CN (16--15)       &  292 -- 296 &   21.4       &  7.7      \\
CH$_3$CN (19--18)       &  329 -- 333 &   19.0       &  6.9      \\
CH$_3$OH (7--6)         &  336 -- 340 &   18.6       &  6.8      \\
SO (8--7)               &  343 -- 347 &   18.2       &  6.6      \\
CH$_3$OCH$_3$ (11--10)  &  347 -- 351 &   17.9       &  6.5      \\
HCOOCH$_3$ (33--32)     &  352 -- 356 &   17.7       &  6.4      \\
\hline 
\end{tabular}
\end{table}
\begin{figure*}[!t]
\centering{
\includegraphics[width=13cm,keepaspectratio]{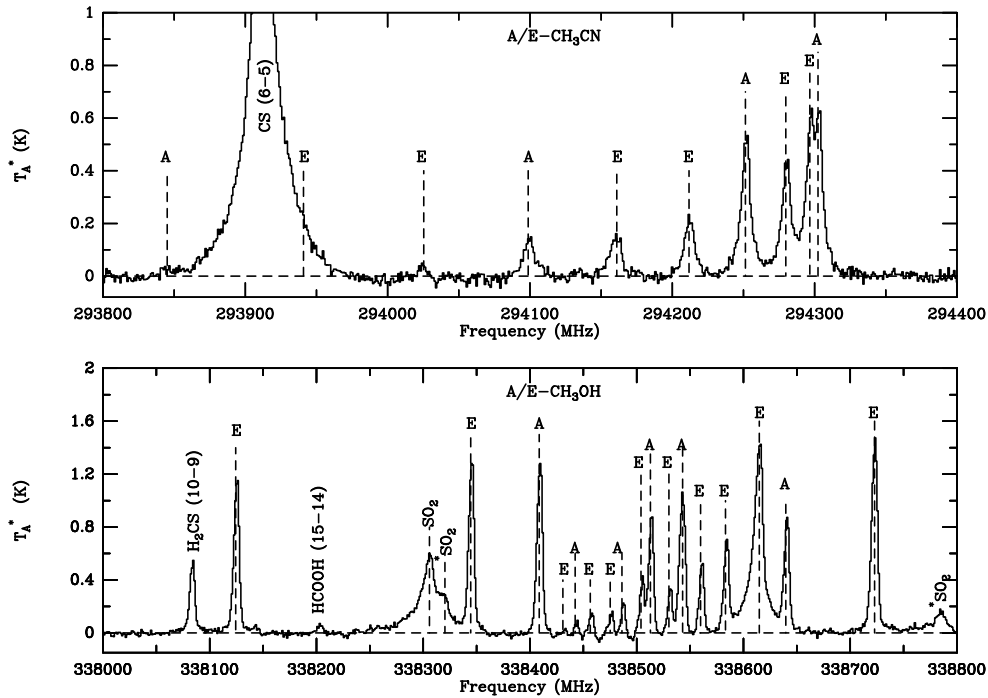}
\caption{Lines of $A/E$-CH$_3$CN and $A/E$-CH$_3$OH as observed with the APEX-2 SHeFI instrument towards G331. Lines were adjusted to their rest frequencies using the CDMS and JPL catalogues. Due to the high intensity of the methanol lines, the range of the $y$-axis is different in both panels.}
\label{fig1}}
\end{figure*}

The data reduction was carried out with the CLASS package of the GILDAS software.\footnote{\url{https://www.iram.fr/IRAMFR/GILDAS/}}
The line identification was performed using the  NIST\footnote{\url{http://physics.nist.gov/cgi-bin/micro/table5/start.pl}} \citep{nist}, CDMS\footnote{\url{https://www.astro.uni-koeln.de/cdms}} \ \citep{muller2005} \ and JPL\footnote{\url{https://spec.jpl.nasa.gov/}} \ \citep{pickett1998} \ spectroscopy databases. Specifically, the CDMS and JPL catalogues were interactively loaded on the survey using the Weeds extension of CLASS. The radiative analyses were performed using the CASSIS  software.\footnote{\url{http://cassis.irap.omp.eu/}}

\section{Results}
\label{sec3}

\subsection{A/E-CH$_3$CN}
\begin{table*}[t!]
\centering
\caption{Spectral line analysis of $A$-CH$_3$CN and $E$-CH$_3$CN in G331. Columns 1--4 list the spectroscopic parameters collected from databases such as CDMS and JPL. Columns 5--7 contain the resulting parameters and uncertainties derived from Gaussian fits, they are line flux ($\int T_{mb} dV$), local standard of rest velocity ($V_{lsr}$) and the line width (FWHM). \label{tab2}}
\resizebox{15cm}{!}{
\begin{tabular}{lllllll}
\hline
\hline
Transition & Frequency  & $E_u$ & $A_{ij}$ & $\int T_{mb} dV$ & $V_{lsr}$    & FWHM   \\
           & (MHz)      & (K)        & (10$^{-3}$ s$^{-1}$) & (K km s$^{-1}$)    &  (km s$^{-1}$) &  (km s$^{-1}$) \\
\hline           
$A$-CH$_3$CN	&		&		&		&		&						\\
\tablenotemark{$\dagger$}16$_9$--15$_9$	&	293845.164	&	697.83	&	1.51	&	--	  	&	--	  	&	--	  	  	\\
16$_6$--15$_6$	&	294098.867	&	377.0721	&	1.9	&	2.065	$\pm$ 0.006	&	-90.7	$\pm$ 0.2	&	10.3	$\pm$ 0.5		\\
16$_3$--15$_3$	&	294251.461	&	184.3532	&	2.13	&	5.9	$\pm$ 0.3	&	-90.4	$\pm$ 0.2	&	8.2	$\pm$ 0.6		\\
16$_0$--15$_0$	&	294302.388	&	120.0696	&	2.21	&	7.9	$\pm$ 0.2	&	-89.69	$\pm$ 0.07	&	8.8	$\pm$ 0.2		\\
\tablenotemark{a}18$_9$--17$_9$	&	330557.569	&	728.67	&	2.36	&	0.15	$\pm$ 0.04	&	-91.2	$\pm$ 0.1	&	0.9	$\pm$ 0.2		\\
\tablenotemark{b}18$_6$--17$_6$	&	330842.762	&	407.9468	&	2.8	&	1.53	$\pm$ 0.07	&	-88.2	$\pm$ 0.4	&	10	$\pm$ 0.1		\\
18$_3$--17$_3$	&	331014.296	&	215.2437	&	3.07	&	6.6	$\pm$ 0.5	&	-90.3	$\pm$ 0.3	&	8.9	$\pm$ 0.8		\\
18$_0$--17$_0$	&	331071.544	&	150.9654	&	3.16	&	9.1	$\pm$ 0.3	&	-89.5	$\pm$ 0.1	&	9.4	$\pm$ 0.3		\\
\tablenotemark{$\dagger$}19$_9$--18$_9$	&	348911.401	&	745.42	&	2.87	&	--	  	&	--	  	&	--	  		\\
19$_6$--18$_6$	&	349212.311	&	424.7065	&	3.34	&	3.01	$\pm$ 0.01	&	-90.4	$\pm$ 0.2	&	11.5	$\pm$ 0.5		\\
\tablenotemark{c}19$_3$--18$_3$	&	349393.297	&	232.012	&	3.63	&	3.66	$\pm$ 0.07	&	-91.7	$\pm$ 0.3	&	3.8	$\pm$ 0.6		\\
19$_0$--18$_0$	&	349453.7	&	167.7368	&	3.72	&	5.5	$\pm$ 0.3	&	-90.5	$\pm$ 0.1	&	6.8	$\pm$ 0.4		\\
\hline 
$E$-CH$_3$CN	&		&		&		&		&						\\
\tablenotemark{d}16$_8$--15$_8$	&	293940.916	&	568.69	&	1.65	&	--		&	 --	&	--			\\
16$_7$--15$_7$	&	294025.496	&	461.7635	&	1.78	&	0.8	$\pm$ 0.1	&	-88.7	$\pm$ 0.6	&	10	$\pm$ 2		\\
16$_5$--15$_5$	&	294161.001	&	290.5519	&	1.99	&	2.14	$\pm$ 0.09	&	-89.2	$\pm$ 0.2	&	9.8	$\pm$ 0.5		\\
16$_4$--15$_4$	&	294211.873	&	226.3073	&	2.07	&	2.93	$\pm$ 0.08	&	-90.2	$\pm$ 0.1	&	9.3	$\pm$ 0.3		\\
16$_2$--15$_2$	&	294279.75	&	140.6138	&	2.17	&	4.19	$\pm$ 0.09	&	-90.79	$\pm$ 0.08	&	7.9	$\pm$ 0.2		\\
16$_1$--15$_1$	&	294296.728	&	119.1843	&	2.2	&	7.74	$\pm$ 0.07	&	-91.87	$\pm$ 0.07	&	9.4	$\pm$ 0.1		\\
\tablenotemark{e}18$_8$--17$_8$	&	330665.206	&	599.54	&	2.52	&	1.21	$\pm$ 0.07	&	-91.4	$\pm$ 0.2	&	8.1	$\pm$ 0.5		\\
18$_7$--17$_7$	&	330760.284	&	492.6304	&	2.67	&	0.8	$\pm$ 0.1	&	-90.8	$\pm$ 0.7	&	11	$\pm$ 2		\\
18$_5$--17$_5$	&	330912.608	&	321.4329	&	2.91	&	2.9	$\pm$ 0.1	&	-89.4	$\pm$ 0.2	&	11.6	$\pm$ 0.6		\\
18$_4$--17$_4$	&	330969.794	&	257.1937	&	3	&	3.3	$\pm$ 0.1	&	-90.5	$\pm$ 0.1	&	9.2	$\pm$ 0.4		\\
18$_2$--17$_2$	&	331046.096	&	171.5073	&	3.12	&	6.66	$\pm$ 0.09	&	-90.87	$\pm$ 0.07	&	10.4	$\pm$ 0.2		\\
18$_1$--17$_1$	&	331065.181	&	150.0797	&	3.15	&	10.252	$\pm$ 0.007	&	-91.62	$\pm$ 0.06	&	10.6	$\pm$ 0.1		\\
\tablenotemark{$\dagger$}19$_8$--18$_8$	&	349024.971	&	616.29	&	3.05	&	--		&	--		&	--			\\
\tablenotemark{f}19$_7$--18$_7$	&	349125.287	&	509.38	&	3.2	&	0.304	$\pm$ 0.003	&	-90.1	$\pm$ 0.6	&	5.9	$\pm$ 0.7		\\
19$_5$--18$_5$	&	349286.006	&	338.1962	&	3.45	&	2.106	$\pm$ 0.008	&	-89.6	$\pm$ 0.2	&	9.5	$\pm$ 0.5		\\
\tablenotemark{g}19$_4$--18$_4$	&	349346.343	&	273.9599	&	3.55	&	2.091	$\pm$ 0.002	&	-89.76	$\pm$ 0.09	&	6.1	$\pm$ 0.1		\\
19$_2$--18$_2$	&	349426.85	&	188.2773	&	3.67	&	5.3	$\pm$ 0.1	&	-90.61	$\pm$ 0.9	&	9.8	$\pm$ 0.2		\\
19$_1$--18$_1$	&	349446.987	&	166.8506	&	3.71	&	8.018	$\pm$ 0.001	&	-91.76	$\pm$ 0.05	&	9.39	$\pm$ 0.08		\\
\hline
\end{tabular}
}
\tablecomments{Likely blended lines with (a) HCOOCH$_2$CH$_3$ (330557~MHz) and/or c-C$_3$HD (330558~MHz), (b) HNCO (330848~MHz), (c) C$_2$H (349400~MHz),  (d) CS (293912~MHz), (e) $^{34}$SO$_2$ (330667~MHz), (f) CH$_3$OCHO (349124~MHz) and (g) C$_2$H (349339~MHz). $\dagger$Marginal emission ($<5\sigma$).} 
\end{table*}
%

\begin{table*}[t!]
\centering
\caption{The table~\ref{tab2} caption applies here for $A$-CH$_3$OH and $E$-CH$_3$OH. \label{tab3}}
\resizebox{15cm}{!}{
\begin{tabular}{llllllll}
\hline
\hline
Transition & Frequency  & $E_u$ & $A_{ij}$ & $\int T_{mb} dV$ & $V_{lsr}$    & FWHM   \\
$J_k$           & (MHz)      & (K)        & (10$^{-5}$ s$^{-1}$) & (K km s$^{-1}$)    &  (km s$^{-1}$) &  (km s$^{-1}$)  \\
\hline
$A$-CH$_3$OH	&		&		&		&		&		&	&				\\
6$_{1}$--5$_{1}$ 	&	292672.889	&	63.707	&	10.6	&	6.95	$\pm$ 0.06	&	-90.67	$\pm$ 0.02	&	5.24	$\pm$ 0.05		\\	
 3$_{2}$--4$_{1}$	&	293464.055	&	51.6381	&	2.89	&	3.39	$\pm$ 0.09	&	-90.35	$\pm$ 0.09	&	6.8	$\pm$ 0.2		\\
 12$_{2}$--11$_{3}$	&	329632.881	&	218.8033	&	6	&	0.59	$\pm$ 0.07	&	-89.8	$\pm$ 0.3	&	4.1	$\pm$ 0.6		\\
 11$_{1}$--11$_{0}$	&	331502.319	&	169.0101	&	19.6	&	3.394	$\pm$ 0.001	&	-90.17	$\pm$ 0.02	&	2.82	$\pm$ 0.03		\\
\tablenotemark{$\dagger$}7$_{3}$--8$_{1}$	&	332920.822	&	114.794	&	1.15 $\times$ 10$^{-5}$ 	&	--		&	--		&	--			\\
 12$_{1}$--12$_{0}$	&	336865.149	&	197.0765	&	20.3	&	6.9	$\pm$ 0.5	&	-91.9	$\pm$ 0.2	&	6.9	$\pm$ 0.3		\\
 7$_{0}$--6$_{0}$	&	338408.698	&	64.9817	&	17	&	10.83	$\pm$ 0.07	&	-90.51	$\pm$ 0.02	&	5.31	$\pm$ 0.04		\\
 7$_{6}$--6$_{6}$	 &	338442.367	&	258.6994	&	4.53	&	1.09	$\pm$ 0.06	&	-91.3	$\pm$ 0.1	&	5.7	$\pm$ 0.4		\\
 7$_{5}$--6$_{5}$	 &	338486.322	&	202.8864	&	8.38	&	2.11	$\pm$ 0.08	&	-90.98	$\pm$ 0.09	&	4.9	$\pm$ 0.2		\\
 7$_{2}$--6$_{2}$	&	338512.853	&	102.7032	&	15.7	&	7.3	$\pm$ 0.4	&	-90.5	$\pm$ 0.1	&	5.5	$\pm$ 0.4		\\
 7$_{3}$--6$_{3}$	&	338543.152	&	114.7948	&	13.9	&	9.21	$\pm$ 0.08	&	-89.77	$\pm$ 0.02	&	5.62	$\pm$ 0.05		\\
 7$_{2}$--6$_{2}$	&	338639.802	&	102.7179	&	15.7	&	7.4	$\pm$ 0.7	&	-90.6	$\pm$ 0.2	&	5.6	$\pm$ 0.6		\\
 2$_{2}$--3$_{1}$	&	340141.143	&	44.6729	&	2.79	&	2.14	$\pm$ 0.06	&	-90.71	$\pm$ 0.08	&	5.5	$\pm$ 0.2		\\
\tablenotemark{$\dagger$}16$_{1}$--15$_{2}$	&	345903.916	&	332.6545	&	9.02	&	--	& --	& --		\\
 5$_{4}$--6$_{3}$	&	346204.271	&	115.1625	&	2.17	&	2.31	$\pm$ 0.07	&	-90.74	$\pm$ 0.09	&	5.9	$\pm$ 0.2		\\
 14$_{1}$--14$_{0}$	&	349106.997	&	260.2068	&	22	&	4.25	$\pm$ 0.06	&	-90.77	$\pm$ 0.04	&	5.41	$\pm$ 0.09		\\
 1$_{1}$--0$_{0}$	&	350905.1	&	16.841	&	33.1	&	12.19	$\pm$ 0.05	&	-90.58	$\pm$ 0.01	&	5.13	$\pm$ 0.03		\\
\tablenotemark{$\dagger$}8$_{2}$--9$_{0}$	&	351685.48	&	121.2915	&	16.2	&	--		&	--		&	--			\\
\tablenotemark{$\dagger$}10$_{4}$--11$_{2}$	&	355279.539	&	207.9929	&	1.22 $\times$ 10$^{-3}$	&	--		&	--		&	--			\\
 13$_{0}$--12$_{1}$	&	355602.945	&	211.0285	&	12.6	&	4.02	$\pm$ 0.04	&	-91.01	$\pm$ 0.02	&	4.76	$\pm$ 0.06		\\
 15$_{1}$--15$_{0}$	&	356007.235	&	295.2675	&	23	&	2.68	$\pm$ 0.05	&	-90.46	$\pm$ 0.05	&	5.8	$\pm$ 0.1		\\
\hline
$E$-CH$_3$OH	&		&		&		&		&		&				\\
8$_{-3}$--9$_{-2}$	&	330793.887	&	138.3774	&	5.38	&	1.96	$\pm$ 0.06	&	-91.03	$\pm$ 0.07	&	4.6	$\pm$ 0.1		\\
16$_{-1}$--15$_{-2}$	&	331220.371	&	312.7333	&	5.23	&	0.55	$\pm$ 0.06	&	-91.4	$\pm$ 0.3	&	5.3	$\pm$ 0.6		\\
3$_{3}$--4$_{2}$	&	337135.853	&	53.7412	&	1.58	&	1.09	$\pm$ 0.06	&	-91.1	$\pm$ 0.1	&	4.7	$\pm$ 0.3		\\
\tablenotemark{$\dagger$}7$_{2}$--6$_{2}$	        &	337671.238	&	456.8285	&	15.5	&	--		&	--		&	--			\\
\tablenotemark{$\dagger$}7$_{5}$--6$_{5}$	        &	337685.248	&	486.0568	&	8.28	&	--		&	--		&	--			\\
\tablenotemark{$\dagger$}7$_{-1}$--6$_{-1}$	        &	337707.568	&	470.3161	&	16.5	&	--		&	--		&	--			\\
7$_{0}$--6$_{0}$	&	338124.488	&	70.1794	&	16.9	&	9.09	$\pm$ 0.06	&	-90.74	$\pm$ 0.01	&	5.01	$\pm$ 0.04		\\
7$_{-1}$--6$_{-1}$	&	338344.588	&	62.6521	&	16.6	&	9.61	$\pm$ 0.08	&	-90.63	$\pm$ 0.02	&	4.87	$\pm$ 0.05		\\
\tablenotemark{$\dagger$}7$_{6}$--6$_{6}$	&	338404.61	&	235.8941	&	4.51	&	--	&	--		&	--		\\
\tablenotemark{$\dagger$}7$_{-6}$--6$_{-6}$	        &	338430.975	&	246.0504	&	4.54	&	--		&	--		&	--			\\
7$_{-5}$--6$_{-5}$	&	338456.536	&	181.1017	&	8.34	&	1.3	$\pm$ 0.4	&	-90.9	$\pm$ 0.7	&	5	$\pm$ 2		\\
7$_{5}$--6$_{5}$	&	338475.226	&	193.1625	&	8.34	&	1.5	$\pm$ 0.6	&	-91	$\pm$ 0.1	&	5	$\pm$ 3		\\
7$_{-4}$--6$_{-4}$	&	338504.065	&	144.9961	&	11.5	&	4	$\pm$ 2	&	-91.3	$\pm$ 0.04	&	6	$\pm$ 1		\\
7$_{4}$--6$_{4}$	&	338530.257	&	153.0934	&	11.5	&	3.1	$\pm$ 0.6	&	-91.1	$\pm$ 0.4	&	5	$\pm$ 1		\\
7$_{-3}$--6$_{-3}$	&	338559.963	&	119.8084	&	14	&	4.5	$\pm$ 0.1	&	-90.93	$\pm$ 0.6	&	5.2	$\pm$ 0.1		\\
7$_{3}$--6$_{3}$	&	338583.216	&	104.8115	&	13.9	&	6.9	$\pm$ 0.2	&	-90.99	$\pm$ 0.08	&	6.6	$\pm$ 0.2		\\
7$_{1}$--6$_{1}$	&	338614.936	&	78.1537	&	17.1	&	21.4	$\pm$ 0.8	&	-89.5	$\pm$ 0.2	&	9.9	$\pm$ 0.5		\\
7$_{-2}$--6$_{-2}$	&	338722.898	&	83.0148	&	15.7	&	12.53	$\pm$ 0.07	&	-90.04	$\pm$ 0.01	&	5.61	$\pm$ 0.03		\\
\tablenotemark{$\dagger$}18$_{2}$--17$_{3}$	        &	344109.039	&	411.5062	&	6.79	&	--		&	--		&	--			\\
\hline
\end{tabular}
}
\tablecomments{$\dagger$Marginal emission ($<5\sigma$).}
\end{table*}
\begin{table*}[t!]
\tabletypesize{\footnotesize}
\centering
\caption{The table~\ref{tab2} caption applies here for $A$-CH$_3$OH	$(\nu_t=1)$ and $E$-CH$_3$OH $(\nu_t=1)$. \label{tab4}}
\begin{tabular}{llllllll}
\hline
\hline
Transition & Frequency  & $E_u$ & $A_{ij}$ & $\int T_{mb} dV$ & $V_{lsr}$    & FWHM   \\
$J_k$           & (MHz)      & (K)        & (10$^{-5}$ s$^{-1}$) & (K km s$^{-1}$)    &  (km s$^{-1}$) &  (km s$^{-1}$)   \\
\hline
$A$-CH$_3$OH	$(\nu_t=1)$&		&		&		&		&		&					\\
\tablenotemark{a}7$_{5}$--6$_{5}$ 	&	337546.048	&	485.4	&	8.13	&	1.047	$\pm$ 0.004	&	-92.8	$\pm$ 0.3	&	9.1	$\pm$ 0.5		\\
\tablenotemark{a}7$_{2}$--6$_{2}$ 	&	337625.679	&	363.5	&	15.5	&	10.2	$\pm$ 0.9	&	-90.5	$\pm$ 0.3	&	8	$\pm$ 1		\\
\tablenotemark{a}7$_{2}$--6$_{2}$ 	&	337635.655	&	363.5	&	15.5	&	1.2	$\pm$ 0.2	&	-92.6	$\pm$ 0.6	&	9	$\pm$ 2		\\
7$_{1}$--6$_{1}$ 	&	337969.414	&	390.1	&	16.6	&	0.39	$\pm$ 0.05	&	-90.8	$\pm$ 0.3	&	5.2	$\pm$ 0.7		\\
\hline
$E$-CH$_3$OH	$(\nu_t=1)$&		&		&		&		&		&					\\
\tablenotemark{a}7$_{4}$--6$_{4}$	&	337581.663	&	428.2	&	11.3	&	--		&	--		&	--			\\
\tablenotemark{a}7$_{-2}$--6$_{-2}$	&	337605.255	&	429.4	&	15.6	&	1.513	$\pm$ 0.003	&	-89.8	$\pm$ 0.3	&	11.7	$\pm$ 0.5		\\
\tablenotemark{a}7$_{0}$--6$_{0}$	&	337643.864	&	365.4	&	16.9	&	1.1	$\pm$ 0.8	&	-90.2	$\pm$ 0.2	&	5.9	$\pm$ 0.5		\\
$^b$10$_{-2}$--11$_{-3}$ 	&	344312.267	&	491.91  	&	17.7	&	--	&	--	&	--		\\
\hline
\end{tabular}
\tablecomments{Transitions and quantum numbers from \citet{XU1997} and references therein. Labels indicate the lines partially and totally blended with  (a) $^{34}$SO 8$_8$--7$_7$ (337580~MHz) and (b) SO 8$_8$--7$_7$ (344310~MHz).}
\end{table*}
We observed the transitions CH$_3$CN $J$=16--15, $J$=18--17 and $J$=19--18; a set of them is displayed in Figure~\ref{fig1}. Within those levels, the  spectral resolution was high enough to separate the components $K$=0, 3, 6 and 9 from $K$=1, 2, 4, 5, 7 and 8,  
which correspond to $A$-CH$_3$CN and $E$-CH$_3$CN, respectively (see Table~\ref{tab2}).

The line identification was performed overlapping the laboratory values
on the observed spectral bands at the rest velocity of the source. A high coincidence was obtained, since the
$K$-ladder structures appeared separately and well centred, at the
source's rest frequency (with $V_{lsr}$=-90 km s$^{-1}$), as predicted by the spectroscopic databases (Figure~\ref{fig1}). This inspection revealed that most of the lines have a Gaussian profile, but only few of them presented broad blue and red wings due to the outflow activity. We applied
Gaussian functions to fit and model the emission. For that, the spectral line processing was carried out using the CLASS package, by means of which we fitted baselines and Gaussian functions for each  processed line.  As a result, we have listed in Table~\ref{tab2} the Gaussian fit parameters and uncertainties of each analysed line; those parameters are the integrated line flux ($\int T_{mb} dV$ in K~km~s$^{-1}$), local standard of rest (LSR) velocity ($V_{lsr}$ in km~s$^{-1}$) and the line width given by the full width at half maximum (FWHM in km~s$^{-1}$).

From a simple comparison of the $A$-CH$_3$CN fluxes, we observed a stiff increase of them as the lines go from the $K$=6 to $K$=0 transitions. This behaviour is in agreement with the probabilities predicted by the Einstein coefficients ($A_{ij}$). For the transitions with $J_{up}$=16, 18 and 19, the obtained patterns are $16_6$:$16_3$:$16_0 \approx$  1:3:4, $18_6$:$18_3$:$18_0 \approx$ 1:4:6 and $19_6$:$19_3$:$19_0 \approx$ 1:1:2, respectively. Only few transitions diverged from this pattern, namely, $A$-CH$_3$CN  $J$=18$_6$--17$_6$ and $J$=19$_3$--18$_3$, whose emission appeared blended with HNCO (15$_{1,14}$-14$_{1,13}$)  and C$_2$H ($4_{4,4}-3_{3,2}$) at 330848~MHz and 349400~MHz, respectively. Those contaminant transitions were previously detected by \citet{jew1989} and \citet{sutt1991} in Orion~A and OMC-1.

For $E-$CH$_3$CN lines, the derived fluxes also increase when the $K$-ladders vary from 7 to 1. That is in agreement with the spontaneous decay rates predicted by $Aij$. For the transitions with $J_{up}$=16, 18 and 19, the complete patterns are
$16_7$:$16_5$:$16_4$:$16_2$:$16_1 \approx$  1:2:3:5:10, $18_7$:$18_5$:$18_4$:$18_2$:$18_1 \approx$ 1:3:4:8:12 and $19_7$:$19_5$:$19_4$:$19_2$:$19_1$ $\approx$ 1:7:7:12:26, respectively.  Here the transition $E$-CH$_3$CN $J$=19$_4$--18$_4$ at 349346~MHz appeared blended with C$_2$H ($4_{5,4}-3_{4,3}$) at 349339~MHz. We also found that the transition $E$-CH$_3$CN $J$=$19_7-18_7$ at 349125~MHz might be blended with CH$_3$OCHO at 349124~MHz.\\

The search for the high levels $K$=8 and $K$=9 yielded negative results, we did not detect lines of these $K$-ladder numbers above the detection limit of the APEX data. For the cases $A$-CH$_3$CN (16$_9$-15$_9$) and (19$_9$-18$_9$), we did not observe emission at all.  In the case of the 18$_9$-17$_9$  line, we  obtained a marginal flux  from the Gaussian fits. This line also might be blended with c-C$_3$HD and/or {\it weeds},\footnote{Conventional term adopted to label species with numerous spectral lines in the mm/sub-mm regime.} see Table~\ref{tab2}. For the cases $E$-CH$_3$CN ($16_8$-$15_8$) and ($18_8$-$17_8$), the emission appeared fully and partially blended with CS (6--5) and $^{34}$SO$_2$ (21$_{2}$-21$_{1}$), respectively. For the $19_8$-$18_8$ transition,  we did not detect emission. The results are listed in Table~\ref{tab2}.  	
In addition, a search for emission of  CH$_3$CN ($v_8=1$) was performed with the current dataset; however we did not observe clear evidences to conclude about its presence in G331.
\subsection{A/E-CH$_3$OH}
As a forest of spectral lines (Figure~\ref{fig1}), the main part of the CH$_3$OH emission has appeared only in an interval of 3~GHz along the whole survey. The lines were identified by a simple comparison of their frequencies, corrected by the source $V_{lsr}$=-90~km~s$^{-1}$, with the rest frequencies reported in molecular databases. The methanol lines exhibit a symmetrical distribution which were fitted by simple Gaussian functions. The spectroscopic and fit parameters estimated for these lines are listed in Table~\ref{tab3}. We find a FWHM of around~5.5~km~s$^{-1}$ and a $V_{lsr}$=-90~km~s$^{-1}$ with a dispersion of $\pm$~2km s$^{-1}$, mainly due to lines which exhibited a low signal-to-noise ratio. In Figure~\ref{fig1} we show a set of $A$ and $E$-CH$_3$OH lines with $J$=7--6.

\begin{figure}[!t]
\centering{
\includegraphics[width=8cm,keepaspectratio]{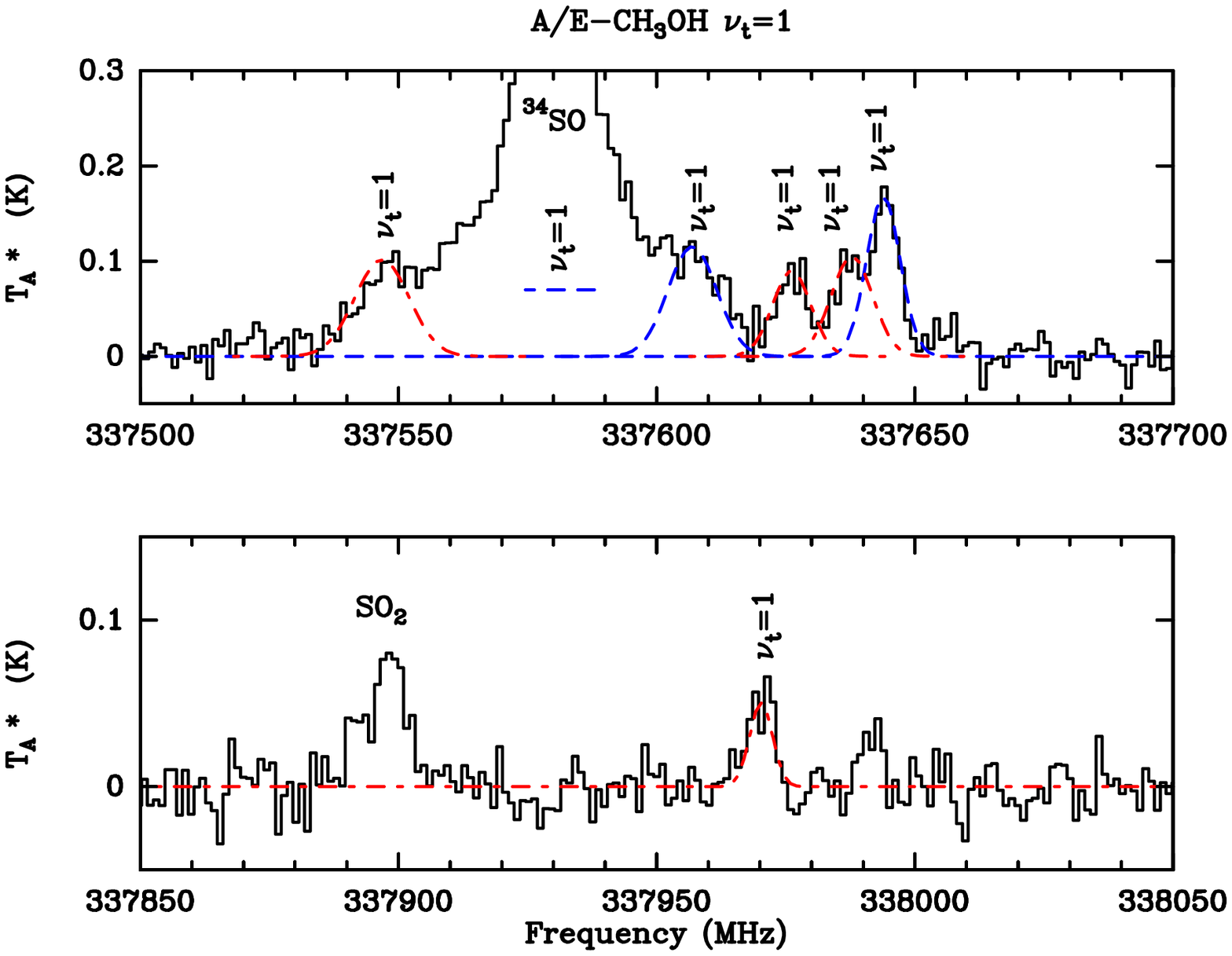}
\caption{Panels showing the torsional excited lines of $A$-CH$_3$OH (red dash-dotted line) and $E$-CH$_3$OH (blue dashed line) against the spectra of G331 (black histogram). In increasing order, the Gaussian fits are centred at the rest frequencies $\sim$~337546~MHz, 337581~MHz (blended with $^{34}$SO), 337605~MHz, 337625~MHz, 337635~MHz, 337643~MHz and 337969~MHz.}
\label{fig2}}
\end{figure}

Although most of the methanol emission corresponds to the torsional ground state, we also identified lines corresponding to the first excited state ($\nu_t$=1). Such is the case for the lines displayed in Figure~\ref{fig2} (upper panel), where it is observed
a broad line of $^{34}$SO at $\sim$~337580~MHz blended with the transition $E$-CH$_3$OH ($\nu_t$=1) $7_4-6_4$ at a frequency, corrected by the source $V_{lsr}$, of~337581.663~MHz. Around the $^{34}$SO peak, minor satellite emission also was identified as corresponding to ($A/E$)-CH$_3$OH $\nu_t=1$; the spectroscopic parameters are summarized in Table~\ref{tab4}. 
In the bottom panel of Figure~\ref{fig2} we display the only excited line without contaminant emission, which corresponds to $A$-CH$_3$OH ($\nu_t$=1) $7_1-6_1$ at a frequency, corrected by the source $V_{lsr}$, of~337969.414~MHz.

Emission of excited methanol is typically observed towards hot molecular cores. For instance, \citet{Menten1986}, \citet{sutton1995}  and \citet{Schilke2001} have reported its emission in OMC-1,  W3(OH) and W51.\\ The observed lines of excited methanol constitute an evidence for its presence in G331. However, in order to solve questions as whether the excited emission is populated or not at the same conditions of the ground state, it is necessary to carry out complementary observations; for instance, at frequencies between 600--720~GHz. These observations could reveal a larger number of CH$_3$OH $\nu_t$=1,2 lines favouring a better estimation on the physical conditions.    
In the bottom panel of Figure~\ref{fig2}, we displayed a broad line close to 337900~MHz which is associated to SO$_2$; however, that line could be also blended with CH$_3$OH ($\nu_t$=2) $J$=7--6 at 337877~MHz.
\subsection{Isotopologues}
\label{sec3.3}
\begin{table*}[t!]
\tabletypesize{\footnotesize}
\centering
\caption{Spectroscopic and observational parameters of the of the $A-^{13}$CH$_3$OH lines analysed in G331. Similarly, the table~\ref{tab2} caption applies here. \label{tab5}}
\begin{tabular}{lllllllll}
\hline
\hline
Transition & Frequency  & $E_u$ & $A_{ij}$ & $\int T_{mb} dV$ & $V_{lsr}$    & FWHM & $^{\dagger}$C/$^{13}$C  \\
$J_k$           & (MHz)      & (K)        & (10$^{-5}$ s$^{-1}$) & (K km s$^{-1}$)    &  (km s$^{-1}$) &  (km s$^{-1}$) &  \\
\hline
7$_{0}$--6$_{0}$ 	&	330252.798	&	63.4	&	15.8	&	0.60	$\pm$ 0.07	& -90.7	$\pm$ 	0.3	& 5.0	$\pm$ 0.6	& 18.1 $\pm$ 0.7\\
1$_{1}$--0$_{0}$ & 350103.118 & 16.8 & 32.9 &  0.69 $\pm$ 0.05 &  -91.2 $\pm$ 0.2 & 4.3 $\pm$ 0.3 &  17.7 $\pm$ 0.1\\
\hline
\end{tabular}
\tablenotetext{}{Labels--- $\dagger$Flux ratios derived by considering the integrated fluxes of $A$-CH$_3$OH (7$_0$--6$_0$) and (1$_0$--0$_0$) listed in Table~\ref{tab3}.}
\end{table*}

We searched for lines of the D, $^{13}$C, $^{15}$N and $^{18}$O isotopologues of CH$_3$CN and CH$_3$OH. As predicted by the spectroscopic databases, several frequencies of those isotopologues fall across the bands; however we only confirm the presence of $^{13}$CH$_3$OH. Emission related to $^{13}$CH$_3$CN $J$=19--18 was also identified; however, the lines are blended with SO $J$=3--2 at 339341~MHz. Those lines were also reported  by \citet{jew1989} in Orion~A.\\
We detected a couple of transitions of $A$-$^{13}$CH$_3$OH. The associated spectroscopic parameters are presented in Table~\ref{tab5}. The lines were identified at the rest frequencies 330252.798~MHz and 350103.118~MHz with signal-to-noise ratios of $S/N \geq 5$. Such detections have been fundamental to analyse the opacity of the detected emission. 
For instance, we estimated flux ratios using lines of $A$-$^{13}$CH$_3$OH and $A$-CH$_3$OH with comparable spectroscopy, namely, using those transitions which have exhibited  similar frequencies, $A_{ij}$ and $E_u$ values. Then, by selecting the $7_0$--$6_0$ and $1_1$--$0_0$ transitions both of $A$-CH$_3$OH and $A$-$^{13}$CH$_3$OH (Tables~\ref{tab3} and \ref{tab5}), whose $A_{ij}$ values are in agreement almost by a factor 1, we obtained the flux ratios C/$^{13}$C=~18.1 $\pm$ 0.7 and C/$^{13}$C=~17.7 $\pm$ 0.1, respectively. Such flux ratios are similar to the values reported in sources towards the Galactic center, where C/$^{13}$C=~20 \citep{Wilson1994,Requena2006}. These ratios imply finite values of opacities, therefore in the subsequent sections we will take into account optical depth corrections to estimate the physical conditions derived from the CH$_3$OH and CH$_3$CN emission.

Two lines of $E$-$^{13}$CH$_3$OH were also identified, although they exhibited contamination. Close to their rest frequencies, lines of $^{34}$SO$_2$ $J$=8--7 and CH$_3$OCH$_3$ $J$=16--15 are candidates for blended emission. These contaminants were also detected towards the Hot Molecular Core G34.3+0.15 \citep{Macdonald1996}.
\begin{table*}[!t]
\centering
\caption{Parameters and physical conditions derived from the statistical equilibrium calculations which have yielded hydrogen densities between 0.7 and 1 $\times$ 10$^7$~cm$^{-3}$. The number of computed lines and the reduced $\chi^2$ values have also been included.\label{tab6}}
\begin{tabular}{l|cc|cc}
\hline
\hline
Parameters & $A$-CH$_3$CN  & $E$-CH$_3$CN & $A$-CH$_3$OH & $E$-CH$_3$OH \\
\hline
Modelled lines & 7 & 13 & 10 & 13 \\
$\chi^2$ min & 5.69 & 7.77 & 6.23 & 4.42 \\
FWHM (km s$^{-1}$) & 5.49 $\pm$ 0.01 & 5.49 $\pm$ 0.01 & 4.9 $\pm$ 0.1 & 4.88 $\pm$ 0.01\\
$N$  (10$^{14}$ cm$^{-2}$) & 4.77 $\pm$ 0.05   &  3.69 $\pm$ 0.04  & 85 $\pm$ 20  & 98 $\pm$ 0.1  \\
$T_k$ (K)     & 142.1 $\pm$ 0.9  & 140.5 $\pm$ 0.3 &  64 $\pm$ 2 & 84.8 $\pm$ 0.1 \\
$\theta$   ($\arcsec$)      & 4.51 $\pm$ 0.01   & 4.53 $\pm$ 0.02 & 5.7 $\pm$ 0.3 & 5.01 $\pm$ 0.04 \\
\hline
\end{tabular}
\end{table*}
\section{Physical conditions}
\label{sec4}

\subsection{Statistical equilibrium calculations}

In this section we present results based on statistical equilibrium models when considering collisions between the analysed molecules and molecular hydrogen. To estimate the kinetic temperatures and column densities, we ran the RADEX code into CASSIS, whose formalism is analogous to the Large Velocity Gradient approximation \citep{vander2007}. RADEX considers the probability that a photon escape in an isothermal and homogeneous medium without large-scale velocity fields. Collision rate data were collected from the LAMDA database.\footnote{Leiden Atomic and Molecular Database LAMDA, http://home.strw.leidenuniv.nl/moldata/}

Conjointly with RADEX/CASSIS, we used the Markov Chain Monte Carlo (MCMC) method to simulate the observed lines from the resulting best solutions, which are given by the minimal $\chi^2$ value. The MCMC algorithm operates on values chosen randomly within a sample defined by a minimum and maximum limit for an N-dimensional parameter space. This characteristic is the most relevant when calculations require as inputs several free parameters. In detriment, the quality of the fit is a monotonically increasing function of the number of iterations introduced in the chain, which in turn affects the \lq\lq computational time\rq\rq \ to achieve convergence (e.g. \citealt{Guan2007,Foreman2013,Tahani2016}).

\begin{figure*}[!t]
\centering{
\includegraphics[width=13cm,keepaspectratio]{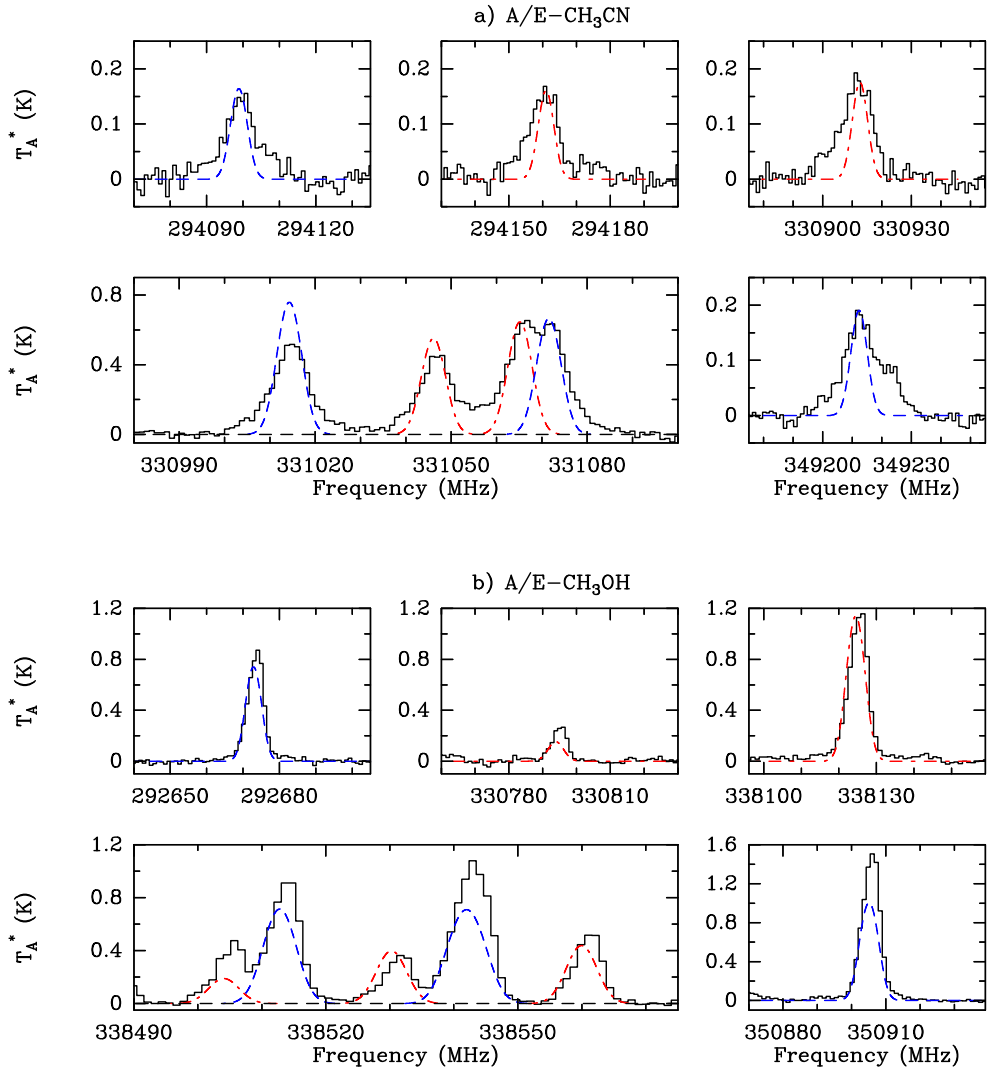}
\caption{Spectral lines and the best fits assuming non-LTE conditions, plotted as a function of the  frequency corrected by the source $V_{lsr}$, of a) $A$-CH$_3$CN (blue dashed-line) and $E$-CH$_3$CN (red dash-dotted-line) and b) $A$-CH$_3$OH (blue dashed-line) and $E$-CH$_3$OH (red dash-dotted-line). For $A/E$-CH$_3$CN, the models yielded a kinetic temperature between $\sim$~(140--142)~K and column densities between $\sim$~(3.7--4.8) $\times$ 10$^{14}$~cm$^{-2}$. For  $A/E$-CH$_3$OH, the models yielded a kinetic temperature between $\sim$~(64--85)~K and column densities between $\sim$~(8.5--9.8) $\times$ 10$^{15}$~cm$^{-2}$.}
\label{fig3}}
\end{figure*}

We prepared MCMC routines for each $A$ and $E$ species. The numerical sample explored by the algorithm was delimited by five free parameters: kinetic temperature ($T_k$), column density ($N$), source size ($\theta$), FWHM and hydrogen density ($n_{\text H_2}$). To guarantee that the code has visited the sample, we tested routines with number of iterations of the order of 10$^5$. Once the algorithm achieves convergence, the minimal $\chi^2$ is computed printing the statistical results for each parameter. This methodology implied large run-times. As a total, we modelled 7 and 13 lines of $A$-CH$_3$CN and $E$-CH$_3$CN, respectively, and 10 and 13 lines of $A$-CH$_3$OH and $E$-CH$_3$OH, respectively. The statistical calculations  are summarized in Table~\ref{tab6}. Likewise, the best fit model lines are exhibited in Figure~\ref{fig3}.\\

There are two aspects which were common for the models of $A/E$-CH$_3$OH and $A/E$-CH$_3$CN. First, we obtained H$_2$ densities in the interval $n_{\text H_2}$ = (0.7--1) $\times$ 10$^7$~cm$^{-3}$, which is in agreement within a factor~$\gtrsim$~1.5 with the density derived from the dust continuum (Herv\'ias-Caimapo et al. {\it submitted}).  Also,
 we intended to differentiate the densities of the ortho-H$_2$ and para-H$_2$ collisional partners (o-H$_2$ and p-H$_2$ respectively). Although studies explain the differences when $A$ and $E$-species collide with o-H$_2$ and p-H$_2$ (e.g. \citealt{Rabli2010}), for our purposes we just conclude that the o-H$_2$/p-H$_2$ ratio varied between $\sim$~0.6 and 1.0 for the different MCMC computations. That interval is consistent with the ortho-to-para limit of 0.2, when H$_2$ is thermalized at the CO temperature, and ortho-to-para limit of  3 when H$_2$ first forms \citep{Flower1984,Lacy1994}. \\
The second aspect is that the modelled spectral lines yielded source sizes in agreement with a compact emitter region. For  CH$_3$CN and CH$_3$OH we found sizes of $\sim$~4.5$\arcsec$ and $\sim$~5.3$\arcsec$, respectively. This result is supported by previous CO (7--6) maps obtained with the Australia Telescope Compact Array (ATCA), for which \citet{bro08} determined a compact and dense  region of $\sim$~4.3$\arcsec$. \citet{mer13b,mer13a} mapping the emission traced by  SiO (8--7) found a ring-type structure of~$\lesssim$~5~$\arcsec$.

\subsection{LTE analysis}
Rotational diagrams were constructed to estimate the excitation
temperatures ($T_{exc}$) and column densities ($N$) of $A/E$-CH$_3$CN and $A/E$-CH$_3$OH. Assuming 
that the emission is optically thin and uniformly fills the 
antenna beam, $T_{exc}$ and $N$ can be calculated from
\begin{equation}\label{eq1}
\ln \left(\frac{N_u}{g_u}\right) = \ln \left(\frac{N}{Z}\right) -\frac{E_u}{kT_{exc}}
\end{equation}
where $N_u/g_u$ and $E_u$ are the column density per statistical weight and
the energy of the upper level, respectively, and $Z$ is the partition function \citep{Goldsmith1999}. In local thermodynamic equilibrium (LTE), Equation~\ref{eq1} represents a Boltzmann distribution whose
values $\ln N_u/g_u$ vs $E_u/k$ can be fitted using a straight line,
whose slope is defined by the term $1/T_{exc}$. 

The rotational diagrams were constructed taking a beam dilution factor associated with a  point-like emitter region. Therefore, a correction that is provided by the term $\ln \left(\Delta \Omega_a/ \Delta \Omega_s\right)$ was introduced on the right-hand of Equation~\ref{eq1} \citep{Goldsmith1999}. This ratio relates the subtended angle of the source with the solid angle of the antenna beam. 
The beam dilution factor that we introduced is based on the results derived from the RADEX/MCMC calculations. Then, for CH$_3$CN and CH$_3$OH we adopted averaged source sizes of 4.5$\arcsec$ and 5.3$\arcsec$, respectively. 

Since the lines observed in this work fell in a broad band, from  290 to 350~GHz,  yielding different beamwidths, the adopted sizes have been used in the comparisons between temperatures and column densities for the different lines.\\

\begin{figure}[!htp]
\centering{
\includegraphics[width=7cm,keepaspectratio]{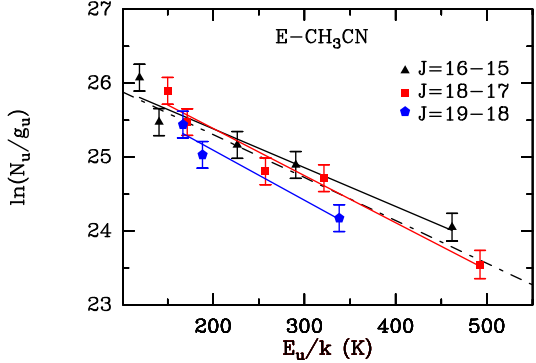}
\caption{Rotational diagram fits of $E$-CH$_3$CN across all the $K$-ladder levels (dashed-dotted line), yielding $N \simeq$ 7.5  $\times$ 10$^{14}$~cm$^{-2}$ and $T_{exc} \simeq$ 172~K, and within each $K$-ladder structure for the transitions $J$=16--15 (black triangles), $J$=18--17 (red squares) and $J$=19--18 (blue pentagons).}
\label{fig5}}
\end{figure}

Depending on the density of the region, the $K$-ladder structures of CH$_3$CN and CH$_3$OH can exhibit an important scattering of~($\ln N_u/g_u,E_u$)-values in their rotational diagrams. More generally, it is expected that the higher is the gas density, the lower is the $K$-ladder segregation    \citep{Olmi1993,Cesaroni1997,Goldsmith1999,Remijan2004,araya2005,cuadrado2017}. In order to check the amount of scatter, we compared the rotational diagrams when they were individually constructed for each $J_K$ level and with those obtained when an unique fit was determined across all the $K$-levels. As a result, we did not observe considerable discrepancies between the two cases, temperatures and column densities were found within the adopted uncertainty from calibration (\S~\ref{sec2}). For instance, in Figure~\ref{fig5} we display the result for $E$-CH$_3$CN.  
\begin{figure*}[!t]
\centering{
\includegraphics[width=12cm,keepaspectratio]{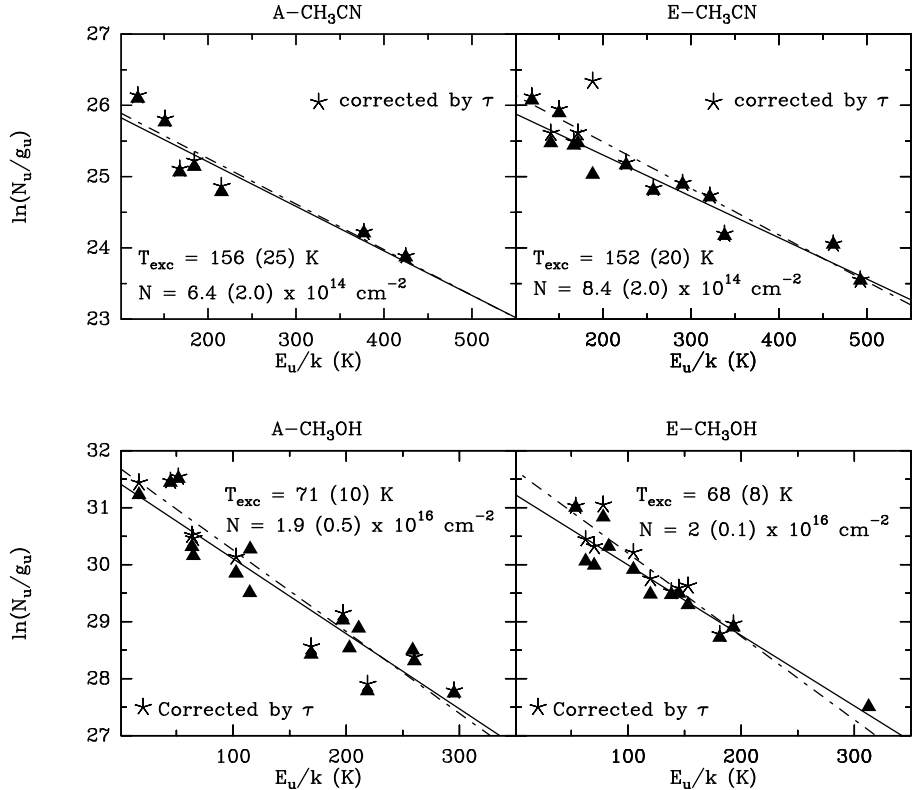}
\caption{Upper panels. Rotational diagrams of $A$-CH$_3$CN and $E$-CH$_3$CN showing the least-square fits with the optical depth correction, represented with the asterisk symbols and dashed-dotted lines, and without the optical depth correction, represented with the filled triangles and  solid lines. Bottom panels, the same as above for $A$-CH$_3$OH and $E$-CH$_3$OH. Column densities and excitation temperatures correspond to the results when applied the optical depth correction, numbers in parenthesis represent uncertainties on the last digit.}
\label{fig4}}
\end{figure*}
\subsubsection*{Optical depth correction}

The rotational diagrams obtained from the $A/E$-CH$_3$CN and $A/E$-CH$_3$OH lines are displayed in Figure~\ref{fig4}. They were constructed selecting only the transitions in the ground state and without contaminant emission. So far, the rotational diagrams have been scaled only considering the beam dilution factor. However, they represent a solution under the assumption of emission optically thin, $\tau \ll 1$. In order to revise how different are the LTE solutions from a radiative scenario with finite values of optical depths,  we have applied an optical depth correction on the rotational diagrams of $A/E$-CH$_3$CN and $A/E$-CH$_3$OH. Expressed mathematically, the correction consisted in including the term $-\ln C_{\tau}$, with $C_{\tau} = \tau / 1 - \exp^{-\tau}$, on the right side of Equation~\ref{eq1} \citep{Goldsmith1999, Gibb2000,Remijan2004,araya2005}. Thus, the ordinates of our rotational diagrams were readjusted considering a term associated with the photon escape probability.\\ 
We have performed this adjust based  on (1$^{st}$) the results obtained from the statistical equilibrium calculations. In the case of CH$_3$CN, transitions with $K$=0,1,2 and 3 were the most affected with $\tau \thickapprox$~1. And, (2$^{nd}$) considering the flux ratio C/$^{13}$C$\lesssim$~20  as we described in section~\ref{sec3.3}, which suggests finite optical depths.
 
In order to perform the optical depth correction, we used CASSIS which executes iterative calculations of $C_{\tau}$ adjusting the level populations until achieving a consistent solution. With these corrections slight changes are expected in the LTE solutions. For instance, \citet{araya2005} obtained differences around 10 and 14~\% in the temperature and column density of CH$_3$CN when the optical depth correction was applied, respectively.

The $A/E$-CH$_3$CN rotational diagrams (RDs), with the optical depth correction, are displayed in the upper panels of Figure~\ref{fig4}. For $A$-CH$_3$CN, we obtained $N$ = (6.4 $\pm$~2.0) $\times$ 10$^{14}$~cm$^{-2}$ and $T_{exc}$ = 156 $\pm$ 25~K.  For $E$-CH$_3$CN, we obtained $N$ = (8.4 $\pm$ 2.0) $\times$ 10$^{14}$~cm$^{-2}$ and $T_{exc}$ = 152 $\pm$ 20~K. The percentage difference of $N$ and $T_{exc}$, with respect to the RD without the optical depth correction, are ranged between (5--14)~\% and (2--12)~\%, respectively.

\begin{figure}[t!]
\centering{
\includegraphics[width=8.5cm,keepaspectratio]{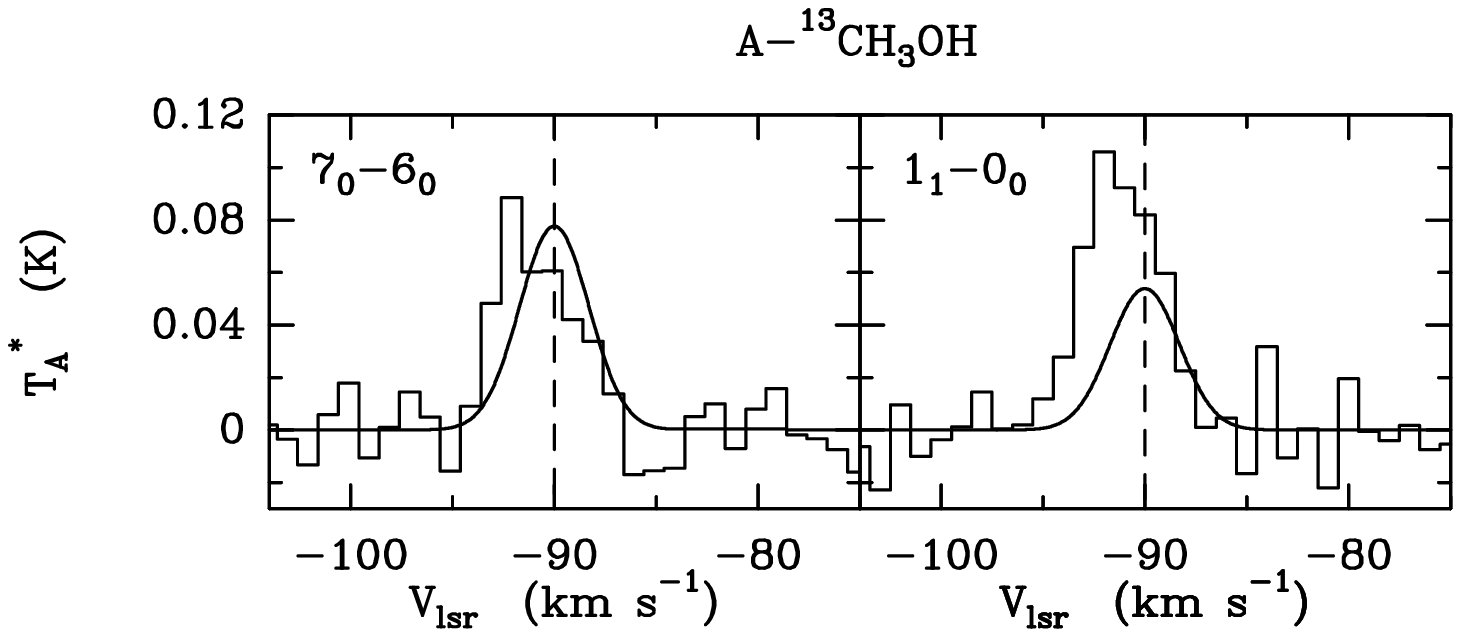}
\caption{Spectral lines  and LTE models of $A$-$^{13}$CH$_3$OH represented by black histograms and Gaussian curves, respectively. Lines $7_0$--$6_0$ and $1_1$--$0_0$ appeared at the rest frequencies 330252.798~MHz and 350103.118~MHz, respectively. The models were derived from a computation whose best solution provided $N$= (1.01 $\pm$ 0.06) $\times$ 10$^{15}$~cm$^{-2}$ and $T_{exc}$= 89.5 $\pm$ 0.4~K.}
\label{fig6}}
\end{figure}

The $A/E$-CH$_3$OH RDs, with the optical depth correction, are displayed in the lower panels of Figure~\ref{fig4}. For $A$-CH$_3$OH, we obtained $N$ = (1.9 $\pm$ 0.5) $\times$ 10$^{16}$~cm$^{-2}$ and $T_{exc}$ = 71 $\pm$ 10~K. For $E$-CH$_3$OH, we obtained $N$ = (2.0 $\pm$ 0.1) $\times$ 10$^{16}$~cm$^{-2}$ and $T_{exc}$ = 68 $\pm$ 8~K. The percentage difference of $N$ and $T_{exc}$, with respect to the RD without the optical depth correction, are ranged between (10--22)~\% and (9--17)~\%, respectively.\\\\
We detected two lines of $^{13}$CH$_3$OH (Table~\ref{tab5}). In the absence of collisional coefficients, those lines were modelled assuming LTE conditions applying the MCMC method leaving various inputs as free parameters: $N$, $T_{exc}$, FWHM, and $\theta$. However, we put an upper limit of $N <$5$\times$10$^{15}$~cm$^{-2}$, which was derived from the column density of the main isotopologue and considering the ratio C/$^{13}$C~$\lesssim$~20. The results were consistent with the physical conditions of the main isotopologue. The best LTE models have been displayed in Figure~\ref{fig6}, whose solution ($\chi^2$=2.82) yielded $N$= (1.01 $\pm$ 0.06) $\times$ 10$^{15}$~cm$^{-2}$ and $T_{exc}$= 89.5 $\pm$ 0.4~K for a modelled source size of 5.7$\arcsec$. Additionally, the resulting FWHM values ($\sim$~4.4~km~s$^{-1}$) are consistent with the resulting fits reported in Table~\ref{tab5}.  
\section{Discussion}
\label{sec5}

\begin{table*}[!t]
\centering
\caption{Abundances of CH$_3$OH and CH$_3$CN derived from the LTE and non-LTE analysis. We have included the contribution of each spin isomer as a percentage of the total abundance, assumed as the sum $A+E$ of the symmetries.\label{tab7}}
\begin{tabular}{l c c c c c c c }
\hline
\hline
Method & \multicolumn{3}{c}{Hot Component} & & \multicolumn{3}{c}{Cold Component}\\
\cline{2-4} \cline{6-8}
  & [CH$_3$CN]   &   [$^{13}$CH$_3$CN] & $\%_{A-\text{CH$_3$CN}}$ &  &  [CH$_3$OH] &   [$^{13}$CH$_3$OH] & $\%_{A-\text{CH$_3$OH}}$  \\
  & 10$^{-9}$ &    10$^{-9}$ &  $\approx$ &  &  10$^{-9}$ &   10$^{-9}$ &  $\approx$ \\
\hline
non-LTE    & 1.8 $\pm$ 0.2  & --   & 56  &  & 38 $\pm$ 10  & --  & 46      \\
LTE        & 3.1 $\pm$ 0.6    & $\lesssim$0.08 & 43  &  & 81 $\pm$ 15 & $\lesssim$2.1 &  49   \\
\hline
\end{tabular}
\end{table*}
\begin{figure*}[t!]
\centering{
\includegraphics[width=17cm,keepaspectratio]{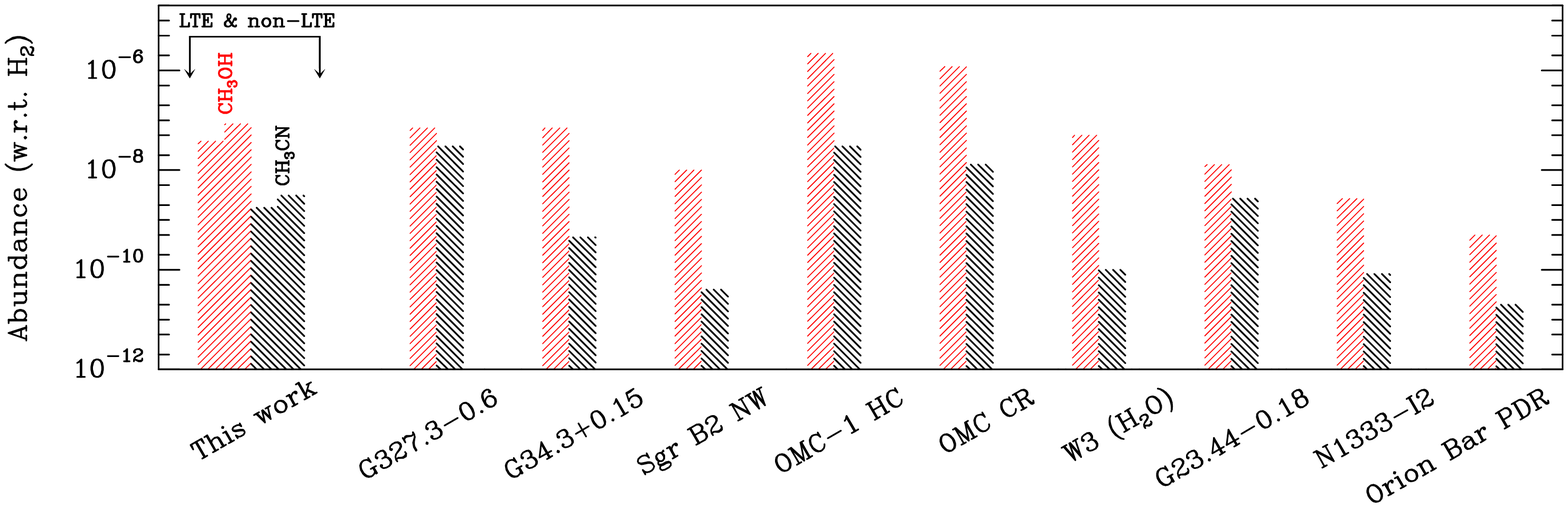}
\caption{Comparison of the abundances of CH$_3$OH (red) and CH$_3$CN (black) in hot molecular cores, clouds, PDR regions. Values were collected from \citet{sutton1995,Macdonald1996,helmich1997,Gibb2000,nummelin2000,Ren2011,crockett2014} and \citet{cuadrado2017}.}
\label{fig7}}
\end{figure*}
\subsection{Molecular abundances}

CH$_3$OH is more abundant and traces a region about 70~K colder than that traced by CH$_3$CN. From the LTE and non-LTE analysis, we found that the ratio ($A+E$)-CH$_3$OH/($A+E$)-CH$_3$CN is about 25, indicating almost the same over abundance of $A$ and $E$-CH$_3$OH over $A$-CH$_3$CN and $E$-CH$_3$CN. \\\\
ALMA observations of the ion H$^{13}$CO$^+$~(4--3) probed the existence of a core and various clumpy structures in G331 (Herv\'ias-Caimapo et~al. {\it submitted}). Maps of SiO~(8--7) show as well an internal cavity surrounded by molecular emission confined in $\lesssim$~5$\arcsec$, for which these authors have estimated $N$(H$^{13}$CO$^+$)$\approx$(1.5--3)~$\times$10$^{13}$~cm$^{-2}$ (e.g. \citealt{bro08,mer13b,mer13a}). 
The column densities of CH$_3$OH and CH$_3$CN were divided by $N$(H$_2$) values to obtain their abundances. Following the works cited above, in order to estimate a scaled density of H$_2$ in G331, we applied the ratio H$^{13}$CO$^+$/H$_2$=3.3 $\times$10$^{-11}$ of Orion KL (e.g. \citealt{Blake1987}) on the H$^{13}$CO$^+$ column densities derived for G331 \citep{mer13b}. As an approximation for the present work, we adopted the H$^{13}$CO$^+$/H$_2$ ratio of the Orion KL system, since this source is an excellent template for comparative studies on abundances and molecular complexity (e.g. \citealt{Schilke2001,Beuther2005}). The total abundances of CH$_3$OH and CH$_3$CN are summarized in Table~\ref{tab7}; as a total, we refer to the $A+E$ contribution from the nuclear spin symmetries (e.g. \citealt{Fuente2014} ). Thus, the CH$_3$CN emission is linked to a hot core with $T_k\approx$141~K and $\theta \approx$4.5$\arcsec$ while CH$_3$OH traces a cold bulk medium with $T_k\approx$74~K and $\theta \approx$5.3$\arcsec$. In the same Table~\ref{tab7}, we also included the abundance uncertainties computed from the errors obtained with the radiative models \citep{Bevington2003}; although we listed errors of $\lesssim$~25\%, the abundance uncertainties could be of up to~40\% including other sources of errors, such as the calibration uncertainty, adopted here as 20\%.
In Table~\ref{tab7}, the abundances of the $^{13}$C isotopologues are also listed. Although we obtained an inaccurate result based only on two lines of $^{13}$CH$_3$OH, we found that it could be not only 20 times lower, as resulting from the $^{13}$C/C flux ratio, but up to 40 times under abundant than CH$_3$OH, as the LTE analysis suggested (see Figure~\ref{fig6}). Thus, we determined [$^{13}$CH$_3$OH]$\lesssim$2.1$\times$10$^{-9}$.\\
For the undetected $^{13}$CH$_3$CN, we propose upper limits assuming that the $^{13}$CH$_3$CN/$^{13}$CH$_3$OH ratio works as what we found for the main isotopologues: CH$_3$CN/CH$_3$OH $\thickapprox$ 1/25. From that conjecture, we conclude that [$^{13}$CH$_3$CN]$\lesssim$8$\times$10$^{-11}$.

One of the goals of this work was to separately analyse the emission of the $A$ and $E$ spin nuclear isomers; however, the analysis indicated common physical conditions ($N$ and $T$) for both spin symmetries meaning that they trace the same reservoir. The opposite case  would imply spin conversion processes, induced by molecular interactions on molecular ices or collisions in gas phase, favouring an over-abundance in a given spin symmetry. However, this result would be expected at earlier stages or in younger stellar sources (e.g. \citealt{Minh1993,Wirstrom2011}).
 
Independently of the employed analysis, we find that both the $A$ and $E$ isomers contribute with approximately half of the total emission of CH$_3$OH and CH$_3$CN. The percentage derived from both methods are listed in Table~\ref{tab7}, as explained above, assuming a total abundance as $A+E$. As a similar result, 
\citet{Macdonald1996} determined the ratio [$A$-CH$_3$OH]/[$E$-CH$_3$OH]$\approx$0.9 for the hot molecular core G34.3+0.15. In the envelopes around low-mass protostars, \citet{Jorgensen2005} found ortho-to-para ratios close to unity for CH$_3$OH. Recently, similar results were reported for the pairs $A/E$-CH$_3$CN, $A/E$-CH$_3$OH and $A/E$-CH$_3$CHO in the Orion Bar photodissociation region (PDR) \citep{cuadrado2017}. \\
This tendency is similar to another aspect aimed to examine in this study: whether o- and p-H$_2$ collisional partners may affect the population of the $A$ and $E$ nuclear spin isomers of CH$_3$OH and CH$_3$CN. However, the MCMC/RADEX computations yielded H$_2$ ortho-to-para ratios close to unity.

Laboratory studies have offered new measurements about spin conversion, generally assumed as improbable processes. In gas phase, \citet{Sun2015} found that molecular collisions induce the interconversion of spin in molecules of methanol exhibiting a rate that decreases as the pressure increases with the number of collisions. A quantum relaxation mechanism explains the interconversion. In condensed phase, \citet{Lee2006} observed that methanol can suffer a slow conversion  from the $E$  to the $A$ symmetry when  it is trapped in a solid matrix of p-H$_2$ prepared at low temperatures ($\sim$~5~K). In theoretical works, \citet{Rabli2010} studied the implications of methanol colliding with o-H$_2$ and p-H$_2$, finding qualitative and quantitative differences for those cases.\\\\
As we summarized in Table~\ref{tab4}, emission of torsionally excited methanol was also evidenced, however only one line appeared without blended emission: $A$-CH$_3$OH ($\nu_t$=1) at $\sim$337969~MHz. In some studies it has been analysed whether excited transitions may be populated at the same LTE conditions of the ground torsional levels
(e.g. \citealt{Lovas82,Menten1986,sutton1995,Ren2011,Sanchez2014}). We made a minor examination based on the unique line without contamination; however, we found that, to account for the flux, different regimes are needed with $T_{exc} \gtrsim$~90~K. Complementary observations with APEX will be carried out to determine with a better precision the physical conditions of excited methanol in G331.

\subsection{The hot and cold components of G331}

In this study we identified two different components that could candidate G331 as a Hot Molecular Core (HMC). Other characteristics also support this hypothesis. For instance, the source is compact, harbours a massive and energetic molecular outflow, and is embedded in a  H II region where masers of OH and CH$_3$OH reveal a high star formation activity. 
Besides, the temperatures of the gas components are above and below the limits where evaporation of icy mantles plays an important role; for instance, to explain abundances of Complex Organic Molecules and aspects related to the age of the source. The gas densities also support such classification, since we determined H$_2$ densities typical of a  dense medium  (0.7--1)~$\times 10^7$~cm$^{-3}$.\\\\ 

N- and O-bearing molecules help to diagnose the presence of different gas components in HMCs (e.g. \citealt{beuther2007,Fontani2007}). In this first work, it is proposed that $A/E$-CH$_3$CN and $A/E$-CH$_3$OH trace a hot ($T_k \simeq$ 141~K) and cold ($T_k \simeq$ 75~K) component with sizes around 4.5$\arcsec$ and 5.3$\arcsec$, respectively. Such results are in agreement with ALMA observations of
CO (7--6), SiO (8--7) and H$^{13}$CO$^+$ (4--3). The emission of these species was found to be compacted in $\sim$~5$\arcsec$ \citep{bro08,mer13b,Hervias2015}.\\

In Figure~\ref{fig7}, we compare the abundances derived in this work with other sources. Our results are similar with the abundances reported in  G34.3+0.15. More generally, our results suggest different abundances, physical components and a chemical differentiation for CH$_3$OH and CH$_3$CN in G331. The high sensitivity and spectral resolution of our single-dish spectra have been useful to realise subtle differences in the line profiles of CH$_3$OH and CH$_3$CN, inspected via the $V_{lsr}$ and FWHM parameters listed in Tables~\ref{tab2} and \ref{tab3}. In agreement with the radiative models, these spectral signatures reveal clues on the origin and size of the emitter regions. In spite of that, and as a perspective, further observations at higher spatial resolution are needed to study the spatial distribution of different tracers in G331. As well, these observations may reveal density gradients towards the core and the outflow, allowing more accurate determinations of molecular abundances.

 The fact of CH$_3$OH being more abundant than CH$_3$CN is in agreement with the chemistry that has been modelled in HMCs, where these molecules are usually referred as parent and daughter species, respectively, since CH$_3$OH is expected to be formed at early stages in grain surfaces, via successive hydrogenation of CO, while CH$_3$CN appears later via reaction between HCN and CH$_3^+$ \citep{Millar1997,Nomura2004,Guzman2013,Loison2014}. Other alternatives to produce methanol are CH$_3^+$ + OH $\rightarrow$ CH$_3$OH and CH$_2$ + H$_2$O $\rightarrow$ CH$_3$OH. They can occur when frozen mixtures of CH$_4$ and H$_2$O are bombarded  with electrons \citep{Hirakoa2006}.
In addition, we have found that the physical conditions traced by those molecules might explain the regimes where pre-biotic and complex organic molecules have been detected, such as CH$_3$OCH$_3$, CH$_3$CHO, NH$_2$CHO and the C$_2$H$_4$O$_2$ isomers.\\ \citet{Nomura2004} performed time-dependent chemical models considering observational aspects of the hot molecular core G34.3+0.15; for instance, that CH$_3$CN is associated to an inner and hot core region while CH$_3$OH traces the gas present in clumps (e.g. \citealt{Hatchell1998,Millar1998,Vandertak2000}). Comparing our column densities with the temporal values calculated by these authors, an age of 10$^4$~yr may represent the epoch for G331, when substantial quantities of daughters molecules (e.g. CH$_3$CN) are produced without substantially decimate their parents (e.g. CH$_3$OH). Preliminary results derived by us, obtained with the gas-grain chemical code Nautilus \citep{ruaud2016},\footnote{http://kida.obs.u-bordeaux1.fr/networks.html} shows a CH$_3$OH/CH$_3$CN ratio similar to those derived from the observations.  However, these results will be presented in a subsequent work depicting the chemistry of O-bearing molecules detected in the source.  

\section{Conclusions and perspectives}
\label{sec6}

As an active interstellar laboratory, the G331.512-0.103 system exhibits a rich chemistry in organic and pre-biotic molecules. In this first article, we analysed around 70 lines of $A/E$-CH$_3$OH and $A/E$-CH$_3$CN towards the central region, abbreviated here as G331. Torsionally excited transitions of methanol were evidenced. Without contaminant emission, we identified the CH$_3$OH ($\nu_t$=1) $7_1$--$6_1$ transition. Likewise, two lines corresponding to $^{13}$CH$_3$OH were detected at 330252.798~MHz and 	350103.118~MHz. 

The analyses were performed including collisions with H$_2$ and typical radiative processes under LTE conditions, namely: rotational diagrams. Both analyses coincide in that CH$_3$OH traces a cold component while CH$_3$CN traces a hotter core with kinetic temperatures of $\sim$~74~K and 141~K, respectively. Likewise, the best-fits indicated emitter regions of around 4.5$\arcsec$ and 5.3$\arcsec$ for these tracers, respectively, with gas density $n$(H$_2$)=(0.7--1) $\times$ 10$^7$~cm$^{-3}$.\\
We treated independently each one of the nuclear spin isomers of CH$_3$OH and CH$_3$CN, and  determined the ratio ($A+E$)-CH$_3$OH/($A+E$)-CH$_3$CN$\simeq$~25. The temperatures and densities of each $A$ and $E$ pair suggest that 
they trace a same bulk and are equally populated at each local temperature. Considering that, we estimated the CH$_3$OH and CH$_3$CN abundances from the total contribution $A+E$. Under the LTE formalism, we estimated that [CH$_3$OH]$\thickapprox$ 8.1 $\times$ 10$^{-8}$, [CH$_3$CN]$\thickapprox$ 3.1$\times$ 10$^{-9}$, and the upper limits  [$^{13}$CH$_3$OH]$\thickapprox$ 2.1 $\times$ 10$^{-9}$ and [$^{13}$CH$_3$CN]$\thickapprox$ 8$\times$ 10$^{-11}$.\\
Under the perspective of hot molecular cores, the CH$_3$OH/CH$_3$CN ratio could be associated with and epoch between (10$^4$--10$^5$)~yr, when daughter molecules like CH$_3$CN start to be produced from parent molecules, such as CH$_3$OH. 

\acknowledgments We thank the anonymous referee for constructive comments and suggestions on the paper. We thank the APEX staff for their helping during the observations. N.U.D. acknowledges support from CONICET, projects PIP 00356  and  from UNLP, projects 11G/120 and PPID/G002. L.B., R.F. and N.R. acknowledge support from CONICYT project BASAL PFB-06. E.M. and J.R.D.L. acknowledge support from the grant 2014/22095-6, S\~ao Paulo Research Foundation (FAPESP).

\vspace{5mm}
\facility{Atacama Pathfinder EXperiment, APEX telescope}

\software{CASSIS (\url{http://cassis.irap.omp.eu/}), GILDAS (\url{https://www.iram.fr/IRAMFR/GILDAS/}), RADEX (\url{http://doi.org/10.1051/0004-6361:20066820})
NAUTILUS (\url{https://doi.org/10.1093/mnras/stw887})
            }  	

\bibliographystyle{aasjournal}
\bibliography{bibliografia-g331}





\end{document}